# Excitations and relaxation dynamics in multiferroic GeV$_4$S$_8$ studied by THz and dielectric spectroscopy

S. Reschke,[1] Zhe Wang,[1,*] F. Mayr,[1] E. Ruff,[1] P. Lunkenheimer,[1] V. Tsurkan,[1,2] and A. Loidl[1,§]

[1]*Experimental Physics V, Center for Electronic Correlations and Magnetism, Institute of Physics, University of Augsburg, 86135 Augsburg, Germany*
[2]*Institute of Applied Physics, Academy of Sciences of Moldova, MD-2028 Chisinau, Republic of Moldova*

We report on THz time-domain spectroscopy on multiferroic GeV$_4$S$_8$, which undergoes orbital ordering at a Jahn-Teller transition at 30.5 K and exhibits antiferromagnetic order below 14.6 K. The THz experiments are complemented by dielectric experiments at audio and radio frequencies. We identify a low-lying excitation close to 15 cm$^{-1}$, which is only weakly temperature dependent and probably corresponds to a molecular excitation within the electronic level scheme of the V$_4$ clusters. In addition, we detect complex temperature-dependent behavior of a low-lying phononic excitation, closely linked to the onset of orbitally-driven ferroelectricity. In the high-temperature cubic phase, which is paramagnetic and orbitally disordered, this excitation is of relaxational character, becomes an overdamped Lorentzian mode in the orbitally ordered phase below the Jahn-Teller transition, and finally appears as well-defined phonon excitation in the antiferromagnetic state. Abrupt changes in real and imaginary parts of the complex dielectric permittivity show that orbital ordering appears via a structural phase transition with strong first-order character and that the onset of antiferromagnetic order is accompanied by significant structural changes, which are of first-order character, too. Dielectric spectroscopy documents that, at low frequencies, significant dipolar relaxations are present in the orbitally ordered, paramagnetic phase only. In contrast to the closely related GaV$_4$S$_8$, this relaxation dynamics that most likely mirrors coupled orbital and polar fluctuations does not seem to be related to the dynamic processes detected in the THz regime.

## I. INTRODUCTION

The cluster compound GeV$_4$S$_8$ belongs to the family of lacunar spinels $AM_4X_8$ ($A$ = Ga and Ge; $M$ = V, Mo, Nb, and Ta; $X$ = S and Se). The structure of lacunar spinels is built by weakly linked cubane ($M_4X_4$)$^{n+}$ and tetrahedral ($AX_4$)$^{n-}$ clusters, which are arranged in a fcc-type network [1,2,3]. The cubane clusters are characterized by a unique electronic distribution, with well-defined total spin of the strongly bonded $M_4$ molecule. In many cases, the cluster orbitals are only partly occupied, leaving unpaired electrons in the highest, often degenerate electronic levels. As a consequence, many of the lacunar spinels are Jahn-Teller (JT) active and, as a function of temperature, the orbital degeneracy is lifted by a symmetry-lowering structural phase transition [4,5]. In lacunar spinels a variety of interesting phenomena, such as pressure-induced superconductivity [6,7], substantial resistive switching [8,9,10], and metal-to-insulator transitions [11,12] were identified. They also belong to the rather rare class of materials, where orbital order, which is established at the JT transition, induces ferroelectricity [13,14]. Ferroelectricity driven by orbital order is a fascinating topic in modern condensed matter physics [15,16,17]. Furthermore, in GaV$_4$S$_8$ a complex low-temperature, low-field - phase diagram has been reported, including cycloidal, skyrmion, and spin-polarized ferromagnetic phases [18,19]. In this compound, in addition to the significant ferroelectric polarization, which arises at the JT transition, all magnetic phases reveal spin-driven excess polarizations, including skyrmions, which are dressed with ferroelectric polarization [14].

---

[*]Present address: Helmholtz-Zentrum Dresden Rossendorf, Bautzner Landstraße 400, 01328 Dresden, Germany
[§]alois.loidl@physik.uni-augsburg.de

Here we report on terahertz (THz) optical and dielectric properties of the semiconducting lacunar spinel GeV$_4$S$_8$, which is a multiferroic compound with orbital-order driven ferroelectricity coexisting with the antiferromagnetic (AFM) ground state. In this compound the highest occupied cluster orbital of the V$_4$ molecules is triply degenerate and filled with two unpaired electrons constituting a total spin $S$ = 1 [3,5]. In GeV$_4$S$_8$ the orbital degeneracy is lifted by a JT-like structural phase transition from the cubic $F\bar{4}3m$ room-temperature phase to the orthorhombic phase with $Imm$2 symmetry at $T_{JT} \approx$ 30-33 K [4,13,20,21]. Orbitally-induced ferroelectricity in GeV$_4$S$_8$ has been clearly identified [13,21] and some experimental evidence has been provided that this structural phase transition consists of two subsequent transitions, which are very close in temperature. In Ref. [21] it was speculated that this sequence of phase transitions could signal the decoupling of ionic and electronic degrees of freedom in establishing ferroelectric order. THz and dielectric spectroscopy is well suited to unravel the coupled orbital and dipolar dynamics. Wang *et al*. [22] systematically studied the relaxation dynamics at the JT transition of the skyrmion host GaV$_4$S$_8$. By combining THz and dielectric experiments at radio frequencies, they found an interesting non-canonical behavior of the polar relaxation dynamics of this material at the orbital-order driven ferroelectric transition. One aim of the present work is to check for similar behavior in the closely related Ge compound.

However, GeV$_4$S$_8$ represents a spin $S$ = 1 system and the resulting JT distortions in the germanium and the gallium compounds are fundamentally different. While in GaV$_4$S$_8$ (S = ½) the V$_4$ tetrahedra are elongated along one out of the four crystallographic <111> directions, in GeV$_4$S$_8$, the V$_4$ tetrahedral units distort with one short and one long vanadium bond on adjacent sites [20]. At lower temperatures, in GeV$_4$S$_8$ antiferromagnetic order is established at $T_N$ = 13 – 18 K [4,5,13,20,21]. In distinct contrast, the skyrmion host GaV$_4$S$_8$ is close to ferromagnetic order and, in zero external magnetic field, reveals cycloidal spin order below 12.7 K [19]. It has to be noted that a THz investigation on ceramic samples of GeV$_4$S$_8$, in a somewhat limited temperature and frequency range when compared to the present study, was very recently published by Warren et al. [23]. Some dielectric results have been published in Refs. [13,21]. However, in this work we mainly focus on the relaxation dynamics. Detailed studies of the temperature dependence of the phonon response of GaV$_4$S$_8$ and GeV$_4$S$_8$ have been presented in Refs. [24,25,26]. While the authors of these studies report on distinct anomalies in the temperature dependence of eigenfrequencies and damping at the JT transition, no indications of a clear soft-phonon behavior, as expected in canonical ferroelectrics, were found. Refs. [24,26,27] provide detailed first-principles calculations of phonon properties for these lacunar spinels.

## II. EXPERIMENTAL DETAILS

Experimental details concerning sample preparation and characterization of GeV$_4$S$_8$ have been given in Ref. [21]. The single crystals, which were used in the THz experiments of this work, have a structural JT transition close to $T_{JT}$ = 30.5 K, where long-range orbital order is induced, and they exhibit antiferromagnetic order below $T_N$ = 14.6 K. Both transitions are strongly of first order [21]. The magnetic susceptibility and magnetization can best be described by localized moments of the $V_4$ molecules with spin $S$ = 1 and strong antiferromagnetic exchange. For transmission THz measurements, a $GeV_4S_8$ single crystal of size 4 × 4 mm$^2$ was prepared, with a thickness of 0.65 mm. The time-domain THz transmission experiments using a TPS Spectra 3000 spectrometer (TeraView Ltd.) were performed perpendicular to the (111) plane for wave numbers from 10 up to 100 cm$^{-1}$ (approximately from 0.3 – 3 THz) in a He-flow cryostat (Oxford Optistat) between liquid-helium and room temperature. In these experiments, transmission and phase shift are measured simultaneously and can directly be converted into a wavenumber-dependent complex permittivity. At low frequencies, the experiments suffer from the limited sample size and from standing waves in the cryostat windows and sample. At high frequencies, the precision is limited by the decreasing intensity of the THz source. The dielectric properties of $GeV_4S_8$ were measured at audio and radio frequencies for temperatures between 4 and 300 K. These experiments in the frequency range from 1 Hz to 10 MHz were performed with a frequency-response analyzer (Novocontrol Alpha-Analyzer). In the dielectric experiments, we used platelet-shaped single crystals of approximately 4 mm$^2$ cross section and 0.4 mm thickness, in most cases with the electric field applied along the crystallographic <111> direction. For this purpose, silver-paint contacts were applied to opposite sides of the crystals.

## III. RESULTS AND DISCUSSION

### A. THz spectroscopy

At THz frequencies, the complex dielectric permittivity $\varepsilon = \varepsilon_1 - i\varepsilon_2$ was determined in the temperature range between 5 and 300 K. Figure 1 shows the temperature dependence of real and imaginary part of $\varepsilon$ for a series of wave numbers between 18 and 78 cm$^{-1}$. Both, the AFM ($T_N$ = 14.6 K) and the structural transition ($T_{JT}$ = 30.5 K) can be identified via significant changes in the temperature and frequency dependences of the dielectric constant $\varepsilon_1$ and loss $\varepsilon_2$. The abrupt changes of these quantities at the phase boundaries document the first-order character of both transitions. The transition temperatures are in perfect agreement with the corresponding values reported in Ref. [21]. Both $\varepsilon_1$ and $\varepsilon_2$ exhibit significant frequency dependences in all three temperature regimes, namely in the magnetically ordered, the orbitally ordered, and the cubic high-temperature phase, which is in contrast to the results in $GaV_4S_8$ [22], where all dispersion effects were nearly suppressed below the JT transition. The dielectric constant $\varepsilon_1$ in $GeV_4S_8$ is about 12 – 13, somewhat lower than in the related gallium compound with $\varepsilon \approx 15$ [22]. It is also worth mentioning that from a recent far-infrared study [25] the dielectric constant at THz frequencies has been determined to be 13.2, in good agreement with the present result.

Figure 1 reveals strong dispersion effects at all temperatures. However, there is no straightforward interpretation of these dielectric THz data. In the paramagnetic and orbitally disordered phase, for 66 and 78 cm$^{-1}$ the temperature dependence of the dielectric loss $\varepsilon_2$ exhibits a well-developed maximum close to 50 K, which slightly shifts to lower temperatures on decreasing frequencies. This signals, at least in this temperature range, relaxational behavior of the dipolar degrees of freedom at the JT transition, similar to the observations in $GaV_4S_8$ [22]. However, the temperature and frequency dependence at lower frequencies indicates that the relaxation dynamics must be more complex. The increase of the dielectric loss towards high temperatures above 200 K most likely signals the importance of dc conductivity contributions at elevated temperatures. These conductivity contributions at high temperatures will not be discussed in the following. Below the transitions, for most frequencies $\varepsilon_2$ is observed to decrease with decreasing temperature. This decrease may signify the left flanks of relaxation or other excitation peaks, whose characteristic times (and, thus, peak temperatures) become shifted at the transitions, preventing the detection of the peak temperature, but clear statements are not possible based on the data shown in Fig. 1 alone.

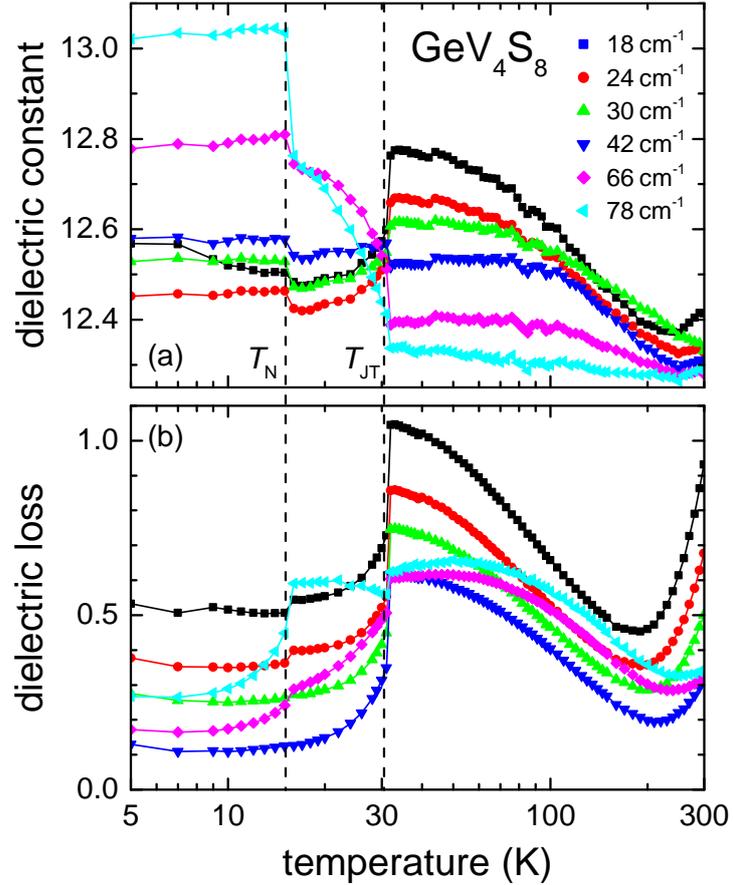

FIG. 1. Semi logarithmic plot of the temperature dependence of the complex dielectric permittivity of GeV$_4$S$_8$. Data are shown between 5 K and room temperature for selected wave numbers between 18 and 78 cm$^{-1}$ in the THz regime (10 cm$^{-1}$ ≈ 0.3 THz). Frames (a) and (b) show the dielectric constant $\varepsilon_1$ and loss $\varepsilon_2$, respectively. The dashed vertical lines indicate the JT transition at $T_{JT}$ = 30.5 K and the onset of AFM order at $T_N$ = 14.6 K. The solid lines connect the data points.

To arrive at a more detailed description of the dynamic processes in GeV$_4$S$_8$, Fig. 2 shows the frequency dependences of the real [Fig. 2(a)] and imaginary part [Fig. 2(b)] of the dielectric permittivity $\varepsilon$, for selected temperatures between 5 and 300 K. In the frequency dependences of both quantities, the phase boundaries close to 15 and 31 K can be easily identified: At these temperatures, the THz response dramatically changes, obviously forced by significant (first order) structural changes, in accord with Fig. 1. Turning first to the imaginary part [Fig. 2(b)], in the complete temperature regime $\varepsilon_2$ exhibits an increase towards low frequencies. At the lowest temperatures, this increase indicates an excitation just at the lower edge of the probed frequency range. When assuming a Lorentzian-type excitation, the decrease of the dielectric constant $\varepsilon_1$ towards low frequencies observed at low temperatures seems to be compatible with this interpretation. Indeed, a well-defined excitation close to 0.3 THz, corresponding to 10 cm$^{-1}$, has been reported by Warren *et al*. [23] between 8 and 24 K. The eigenfrequency of this excitation was found to exhibit almost no temperature dependence, even when crossing the antiferromagnetic phase transition, and has been interpreted as phonon-like excitation in the orbitally ordered phase. We tend to ascribe this excitation to a characteristic mode of the molecular V$_4$ clusters. We would like to recall that a threefold degenerate level forms the highest occupied orbital of these clusters [3]. It could well be that slight local distortions induce a lifting of this degeneracy. The eigenfrequency of approximately 10 cm$^{-1}$ seems to be too low for a canonical phonon-like excitation. Assuming specific symmetry conditions of the high-temperature paraelectric phase, infrared-active phonon modes in GeV$_4$S$_8$ are not expected below 150 cm$^{-1}$, while in the orbitally ordered phase the lowest eigenfrequency was predicted to occur at 67.4 cm$^{-1}$ [26]. Hlinka *et al*. [24] calculated the eigenfrequencies of phonon excitations in the cubic high-temperature phase of GaV$_4$S$_8$ and found the lowest infra-red active mode at 114 cm$^{-1}$.

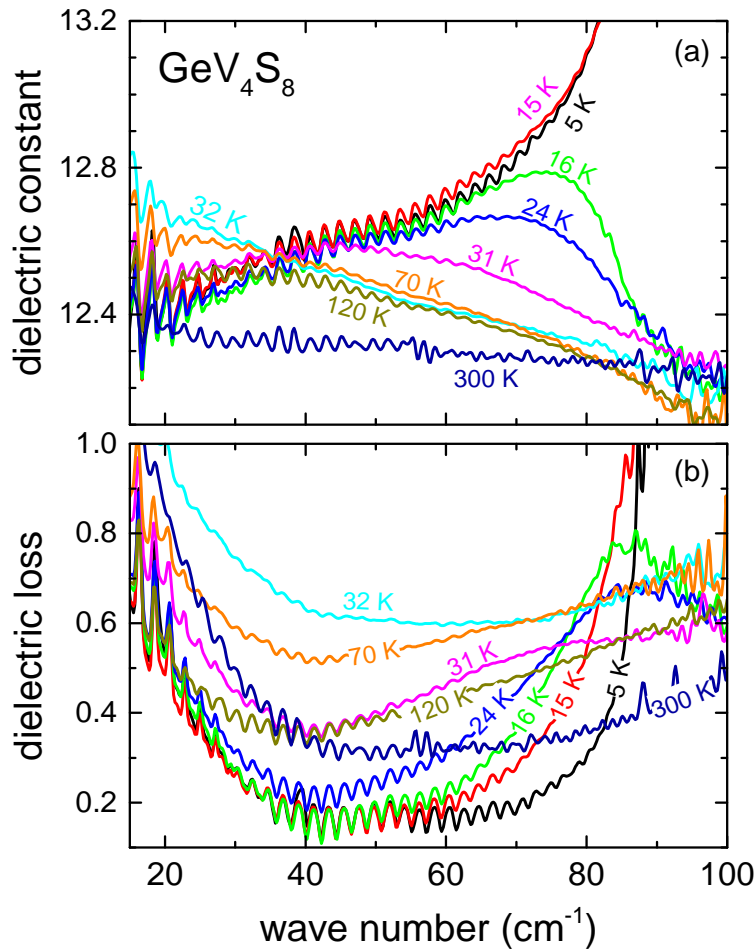

FIG. 2. Frequency dependence of the dielectric constant $\varepsilon_1$ (a) and dielectric loss $\varepsilon_2$ (b) of GeV$_4$S$_8$ between 15 and 100 cm$^{-1}$ for selected temperatures between 5 and 300 K.

When we follow the temperature evolution of this low-lying excitation as documented in Fig. 2(b), on increasing temperatures the increase in $\varepsilon_2$ becomes weaker, seems to be weakest close to 120 K, but then increases again on further increasing temperatures. In the high temperature regime ($T > 120$ K) this increase in $\varepsilon_2$ certainly stems from dc conductivity contributions and can be described by a frequency dependence $\varepsilon_{2,dc} \propto \sigma_{dc}/\omega$.

At the lowest temperatures, in the antiferromagnetically ordered phase, both $\varepsilon_1$ and $\varepsilon_2$ show a strong increase towards higher frequencies. This can naturally be explained assuming a relatively strong Lorentzian-like excitation close to 100 cm$^{-1}$, most probably a phonon excitation of the low-temperature antiferromagnetic phase. Focusing on the frequency dependence of real and imaginary parts of the dielectric permittivity, this phonon excitation abruptly becomes heavily damped for $T > T_N$, in the paramagnetic, but orbitally ordered phase. This is impressively documented in both frames of Fig. 2, when comparing dielectric constant and dielectric loss measured at 5 and 15 K, with values derived at 16, 24, and 31 K. It seems natural to assume that just above $T_N$ orbital fluctuations set in, which strongly couple to this phonon mode. Despite strong damping, the excitations in the orbitally-ordered phase still are of Lorentzian-type with well-defined eigenfrequency. This is best seen by the decrease of the dielectric constant below the maximum in $\varepsilon_1(T)$. However, with further increasing temperatures ($T > 31$ K), this excitation becomes even stronger damped and appears as relaxational mode above the JT transition, in the high-temperature cubic and orbitally disordered phase. In this temperature regime, the coupling of orbital fluctuations to the phonon excitation seems to be even stronger and the THz response probably has to be interpreted in terms of a dynamic JT effect [26]. It is unclear at present, if a small phonon contribution remains at high frequencies close to 100 cm$^{-1}$ in the high-temperature cubic phase.

To describe the frequency-dependent permittivity data in GeV$_4$S$_8$ as documented in Fig. 2, we choose the following approach:

$$\varepsilon = \sum_{j=1}^{3} \frac{\Delta\varepsilon_j \omega_{0,j}^2}{\omega_{0,j}^2 - \omega^2 - i\gamma_j\omega} + \frac{\Delta\varepsilon_D}{1+i\omega\tau} - i\frac{\sigma_{dc}}{\varepsilon_0 \omega} + \varepsilon_\infty \qquad (1)$$

The first term of Eq. (1) describes a sum of Lorentzian oscillators. Here $\Delta\varepsilon$ is the dielectric strength of each mode, $\omega_0$ its (transverse) eigenfrequency and $\gamma$ the damping. The single oscillators are denoted with the indices $j = 1$, 2, and 3, respectively. The second term in Eq. (1) corresponds to a Debye relaxation with the relaxation strength $\Delta\varepsilon_D$ and the mean relaxation time $\tau$. This term is only included for temperatures $T > T_{JT}$ and replaces the Lorentzian contribution of the high-frequency phonon. The third term describes a contribution from dc conductivity $\sigma_{dc}$. Conductivity contributions are relevant only for temperatures $T > 150$ K and are not further considered here.

Three Lorentz oscillators [first term in Eq. (1) with $j = 1$, 2, and 3] are needed to describe the low-frequency excitations in the magnetically ordered phase ($T < T_N$). These include the low- and high-frequency modes discussed above, appearing close to 15 and 95 cm$^{-1}$. As will be documented below, a further excitation at about 55 cm$^{-1}$, which only exists in the orbitally ordered phases, has to be taken into account to arrive at a good description of the experimental results at the lowest temperatures. Three Lorentzian-type oscillators are also needed to describe the spectra in the paramagnetic orbitally ordered phase ($T_N < T < T_{JT}$), with a very low dipolar weight close to experimental uncertainty of the middle excitation at about 55 cm$^{-1}$. This is in accord with the results by Warren *et al*. [23] and this observation discards this excitation as antiferromagnetic resonance. In the high-temperature cubic phase ($T > T_{JT}$) only a single Lorentzian excitation plus a relaxational mode are necessary to describe the THz spectra. A further very weak mode, close to 20 cm$^{-1}$ was detected in Ref. [23] in the AFM phase. We were not able to locate this excitation in our spectra.

Typical results of the fits to the imaginary part of the dielectric permittivity using Eq. (1) are shown in Fig. 3 for selected temperatures (lines). Reasonable agreements of fit and experimental curves are achieved in all phases, namely in the magnetically ordered (5 K), in the paramagnetic and orbitally ordered (16 and 31 K), as well as in the high-temperature cubic phase (32 and 70 K). The significant excitation close to 15 cm$^{-1}$ shows only little temperature dependence. It is slightly shifted and moderately broadened, but can be identified as well-defined excitation in the complete temperature range shown in Fig. 3. Hence, it also seems to exist even in the high-temperature cubic phase. The temperature evolution towards room temperature can hardly be followed, as on increasing temperatures the dc conductivity contribution becomes more important and masks this excitation [see Fig. 2(b)]. In the magnetically ordered phase, a well-defined phonon mode close to 95 cm$^{-1}$ dominates the spectrum. As discussed above, at the transition to the paramagnetic phase, the damping of this phonon excitation abruptly increases and this excitation considerably broadens on further increasing temperatures. However, up to the JT transition at 31 K, despite the unusual broadening, this excitation remains of Lorentzian-type, i.e. with well-defined eigenfrequency. However, in the high-temperature cubic phase, with the loss of orbital order, this excitation becomes fully overdamped and can now be described using a single Debye relaxation, only characterized by a mean relaxation time. The significant change of character of this excitation from Lorentzian to relaxational type is best documented comparing the frequency dependence of the dielectric loss at 31 and 32 K (Fig. 3).

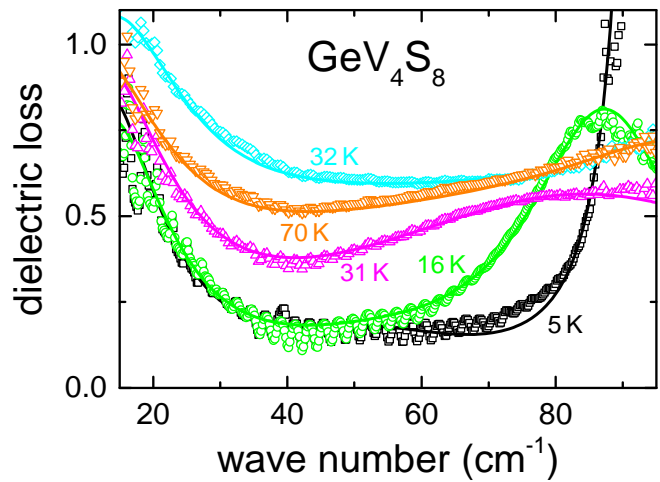

FIG. 3. Frequency dependence of the dielectric loss $\varepsilon_2$ of GeV$_4$S$_8$ (symbols) and corresponding fits using Eq. (1) (solid lines). Spectra are shown for a series of representative temperatures in the AFM and orbitally-ordered phase (5 K), in the paramagnetic and orbitally-ordered phase (16 and 31 K), as well as in the cubic high-temperature phase (32 and 70 K).

Figure 4 shows the temperature dependence of the parameters of the three excitations that dominate the THz response of GeV$_4$S$_8$ below 100 K. At higher temperatures, the spectra are governed by contributions from dc conductivity and the analysis of low-lying excitations becomes rather unreliable. The lowest excitation [Figs. 4(a)-(c)] appears close to 16 cm$^{-1}$ in the AFM phase. As we only were able to measure the high-frequency flank of this mode, we partly fixed the parameters at low temperatures. Overall, our results are in reasonable agreement with those derived by Warren *et al.* [23]. The eigenfrequency shown in Fig. 4(a) exhibits an upward jump to about 19 cm$^{-1}$ just above the JT transition and then decreases on further increasing temperatures. The mode strength [Fig. 4(b)] remains essentially constant up to 100 K within the experimental uncertainties. The damping shows a continuous increase with temperature in the paramagnetic, orbitally ordered phase [Fig. 4(c)]. As mentioned above, the eigenfrequency seems to be too low for a canonical phonon mode. Most probably, this excitation corresponds to a collective mode belonging to the $V_4$ cluster molecules.

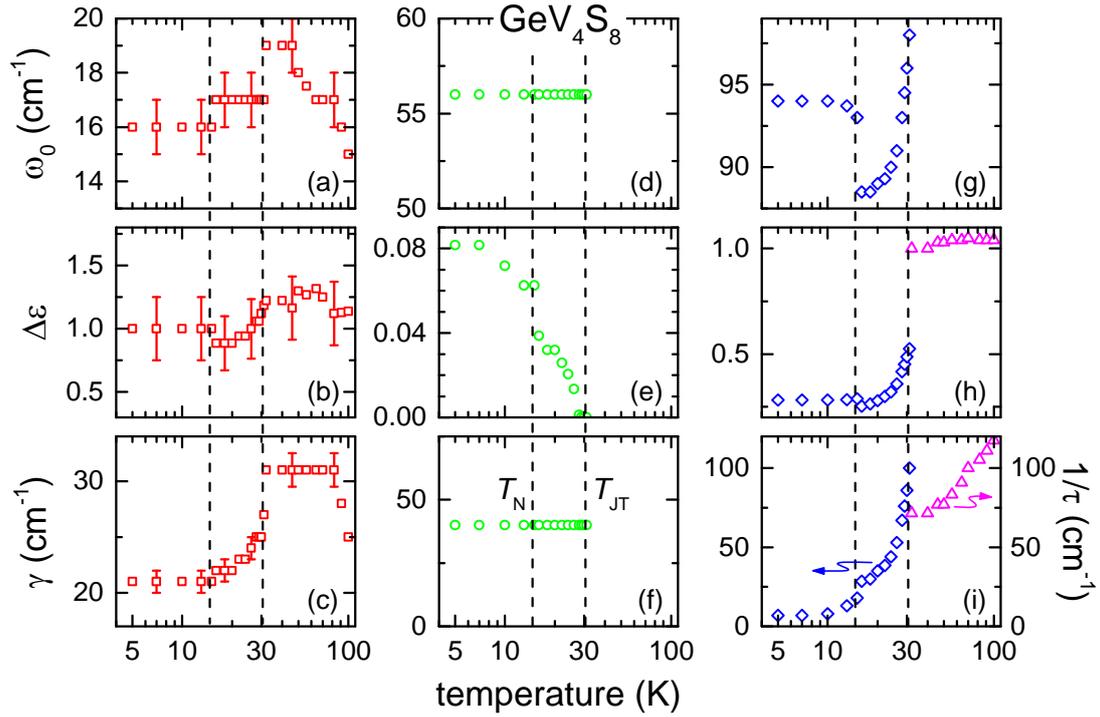

FIG. 4. Temperature dependence of the fitting parameters of the THz response in GeV$_4$S$_8$ for the three modes as documented in Fig. 3 for temperatures $T < 100$ K. The left panels (a – c) show the parametrization of the low-frequency excitation: (a) eigenfrequency, (b) dielectric strength, and (c) damping. The middle panels (d – f) document the temperature evolution of the parameters for the weak resonance close to 55 cm$^{-1}$: (d) eigenfrequency, (e) dielectric strength, and (f) damping. The right panels (g – i) show the parameters observed for the Lorentzian-like phonon mode for $T < T_{JT}$: (g) eigenfrequency, (h) dielectric strength, and (i) damping. For $T > T_{JT}$ a parameterization in terms of a Debye relaxation was used: (h) relaxation strength and (i) mean inverse relaxation time (right axis).

The excitation close to 55 cm$^{-1}$ [Figs. 4(d)-(f)] seems to exist in the orbitally ordered phases only. As it survives the transition into the paramagnetic phase, a magnetic dipole transition probably can be excluded. Due to its very low dipolar strength, for the fits the eigenfrequency and damping had to be fixed. This excitation continuously loses dipolar weight when approaching the JT transition [Fig. 4(e)] and is completely absent in the cubic high-temperature phase. Hence, it seems to be correlated with the onset of orbital order. The fact that its intensity continuously decreases towards the structural phase transition, makes orbital excitations within an excitation scheme due to lifted orbital degeneracy likely. Detailed model calculations will be necessary to arrive at finite conclusions.

Finally, Figs. 4(g)-(i) show the parameters of the strong and dominating phonon mode observed in the AFM phase close to 95 cm$^{-1}$. In the magnetically ordered phase, this phonon mode remains rather constant, with moderately increasing damping on increasing temperatures [Fig. 4(i)]. At the AFM to paramagnetic phase transition, this mode abruptly jumps to lower eigenfrequencies [Fig. 4(g)], but then eigenfrequency and oscillator strength [Fig. 4(h)] continuously increase on further increasing temperatures. Most remarkable, its damping increases by a factor of 4, when the temperature is raised from the AFM to the structural phase transition. Finally, in the high-temperature cubic phase this mode loses its oscillatory character and transforms into a purely relaxational mode.

## B. Dielectric Results

Preliminary results of dielectric measurements at radio and audio-frequencies have been reported previously [13], partly also by our group [21]. Here we focus on the dynamics of relaxational processes and try to compare the results with those derived by THz spectroscopy discussed above. Figure 5 shows the dielectric constant $\varepsilon'(T)$ of GeV$_4$S$_8$ measured at frequencies between 178 Hz and 316 kHz along the crystallographic <111> direction for temperatures from 10 to 50 K. This temperature regime covers the onset of orbital order at 30.5 K and the phase transition into long-range antiferromagnetic order at 14.6 K. The shown data were measured on heating and, for clarity, the dielectric data are shown on a semi-logarithmic plot. Both phase transitions are indicated by abrupt step-like changes of the dielectric constant, which amount more than 100 at low frequencies. These drastic changes of $\varepsilon'$ are not due to Maxwell-Wagner contributions from the surface or from domain walls or grain boundaries [28] but certainly result from abrupt changes in the dipolar relaxation dynamics. The temperature dependence of the dielectric constant reveals a dome-like enhancement in the temperature regime between the two phase transitions. On cooling, the temperature dependence of the dielectric constant is of similar shape (not shown), but exhibits significantly lower values of $\varepsilon'(T)$ in between the structural and magnetic phase transitions. Singh *et al*. [13] have reported very similar behavior of the temperature dependence of the dielectric constant. We would like to pinpoint the fact, that in Ref. [21] the V-shaped minimum in $\varepsilon'(T)$, just below the structural phase transition, was interpreted as possible fingerprint of a separation of electronic (charge distribution) and ionic displacements in the ferroelectric ordering process.

These highly unusual temperature and frequency dependences of the dielectric constant look drastically different than the dispersion effects observed in the THz regime [see Fig. 1(a)]. At first glance, this dielectric response at audio and radio frequencies may be explained by assuming that dipolar relaxations, which dominate the dielectric response in the THz regime in the orbitally disordered phase, are shifted to much lower frequencies at the onset of orbital order, but become fully suppressed again at the magnetic phase transition. An extreme slowing down of relaxation dynamics when orbital order sets in, also characterizes the orbitally-driven ferroelectric transition in GaV$_4$S$_8$ [22]. In this compound, a single dipolar relaxation dominates the THz response in the high-temperature phase, while in the orbitally-ordered phase it shows up at radio frequencies and persists down to the lowest temperatures. This behavior was discussed in terms of an order-disorder type polar transition that is of strongly first order [22]. However, already a first inspection of Figs. 2 and 5 makes clear that the dipolar dynamics at the structural and magnetic phase transitions in GeV$_4$S$_8$ is more complex: For this material, various dynamic processes are observed in the THz regime for all temperatures (Figs. 1 and 2). At audio and radio frequencies, a significant dipolar relaxation process can only be identified in the orbitally-ordered paramagnetic phase (Fig. 5). Just as for GaV$_4$S$_8$, in GeV$_4$S$_8$ we indeed also observe a relaxational process above the JT transition at THz frequencies. However, in contrast to the Ga system [22], it seems unlikely that this THz relaxation reflects the coupled polar and orbital dynamics associated with the ferroelectric transition and shifts to the radio-frequency regime below the JT transition. As discussed above, below $T_{JT}$ it instead remains located in the THz regime and changes its character from relaxational to resonance-like. Thus, the relaxational dynamics that appears at low frequencies in the polar ordered, paramagnetic phase, dominating the dielectric properties at audio/radio-frequencies (Fig. 5), does not seem to be simply connected with the relaxation dynamics in the THz regime, which changes its nature and character but is observable in all phases.

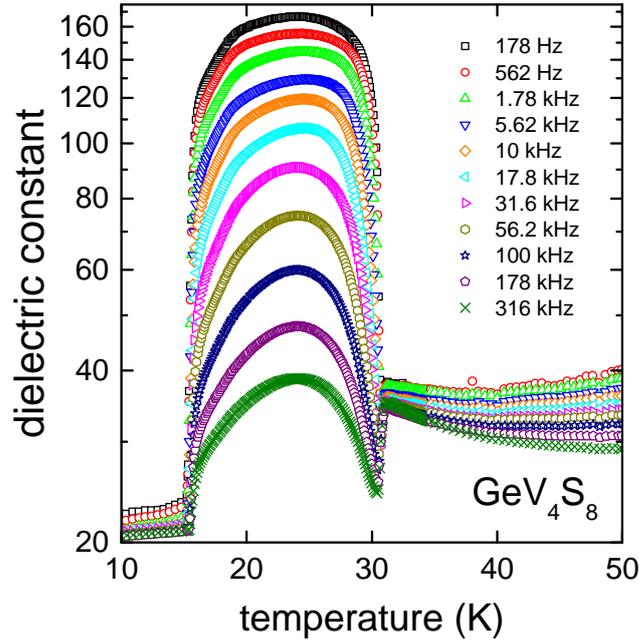

FIG. 5. Temperature dependence of the real part of the permittivity of GeV$_4$S$_8$ measured along the crystallographic <111> direction at frequencies between 178 Hz and 316 kHz under heating on a semi-logarithmic plot.

To affirm the relaxational character of the observed dispersion at audio and radio frequencies and to get an estimate of the temperature evolution of the mean relaxation times, Fig. 6 shows the frequency dependence of the real part of the dielectric permittivity for selected temperatures in the regime between the two phase transitions. Just above the orbital-order transition (35 K) and just below the AFM transition (15 K), the dielectric constant is low, nearly frequency independent, and a significant time scale of a possible relaxation can hardly be determined. However, strong dispersion effects appear between 17.5 and 30 K, just in the orbitally ordered and paramagnetic phase, with a step-like decrease of the dielectric constant from a plateau with high values at low frequencies to low values at high frequencies. This indeed is the typical signature of a relaxational process. The data can reasonably well be fitted assuming a pure mono-dispersive Debye relaxation. The corresponding formula is indicated by the second term in Eq. (1) and the relevant fit parameters are the dielectric strength $\Delta\varepsilon_D$ and the mean relaxation time $\tau$. The fit results are shown as solid lines in Fig. 6 and the temperature dependence of the resulting mean relaxation times is plotted in the inset of Fig. 6. They slow down towards the onset of magnetic order at low and towards the structural phase transition at high temperatures. Between the two phase transitions, the relaxation times are of the order of µs, several decades slower than the relaxation times in the ns range reported for GaV$_4$S$_8$ below $T_{JT}$ [22]. This further supports the notion that, in contrast to the latter compound, for GeV$_4$S$_8$ there is no connection of the relaxational responses detected by the dielectric and THz experiments.

For canonical order-disorder ferroelectrics, by dielectric spectroscopy usually a relaxation process is detected arising from fluctuations of the same dipolar degrees of freedom that lead to polar order [29]. Its temperature-dependent relaxation time rises at the ferroelectric transition. Thus, the detected increase of $\tau(T)$ when approaching the transitions from within the orbitally-ordered paramagnetic phase, documented in the inset of Fig. 6, may well reflect such a behavior. As in GeV$_4$S$_8$, just as for GaV$_4$S$_8$ [22], polar and orbital degrees of freedom seem to be closely coupled, this relaxation process most likely simultaneously mirrors the polar and orbital dynamics, located in the audio and radio frequency range for temperatures between the two transitions.

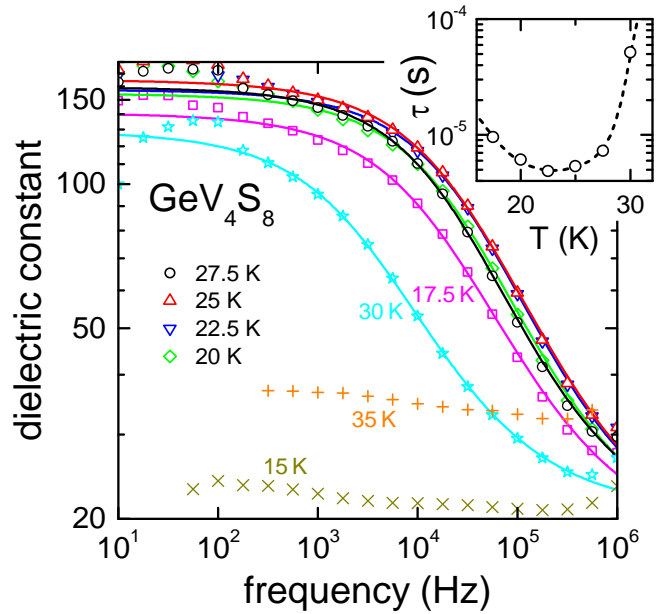

FIG. 6. Dielectric constant *vs*. frequency in GeV$_4$S$_8$ for a series of temperatures between 15 and 35 K on a double-logarithmic plot. The solid lines represent results of mono-dispersive Debye fits as described in the text. The inset shows the temperature dependence of the mean dipolar relaxation time as determined from these fits. The dashed line in the inset is drawn to guide the eye.

## IV. CONCLUSIONS

In this work, by utilizing THz and audio- to radio-wave dielectric spectroscopy we have investigated the polar relaxational dynamics in GeV$_4$S$_8$ associated with the orbital-order driven ferroelectric transition. Our aim was to check for similar non-canonical behavior as found for the order-disorder scenario of the ferroelectric transition in GaV$_4$S$_8$. The main conclusion is that the relaxation dynamics in the germanium compound is completely different when compared to the gallium lacunar spinel. This probably results from the fact that the JT transition in the *S* = 1 germanium system is of different nature and induces a different low-temperature structure, finally resulting in low-temperature antiferromagnetic spin order.

In the THz study, we detected a number of low-lying excitations: Close to 16 cm$^{-1}$, we identified a Lorentzian-type excitation with weak temperature dependence of eigenfrequency, damping, and dielectric strength. This excitation probably corresponds to a transition within the split orbital ground state of the vanadium molecules in the low-symmetry low-temperature phases. As there are still some ambiguities concerning the low-temperature structure of the AFM phase and as no realistic calculations of the orbital scheme of the molecular V$_4$ entities are available, we cannot arrive at finite conclusions. We identify a further weak excitation close to 55 cm$^{-1}$, whose dipolar strength goes to zero, when approaching the JT transition. This low-energy excitation could correspond to a phonon-like excitation as proposed by Warren *et al*. [23]. In the ferroelectric *Imm*2 phase, the lowest IR active phonon frequency with reasonable strength was predicted at 67.4 cm$^{-1}$ [26].

In addition, in the low-temperature magnetically ordered phase most unexpectedly a phonon-like excitation close to 95 cm$^{-1}$ was identified, which reveals a very unusual temperature dependence of both, eigenfrequency and damping. The damping heavily increases (by more than a factor of 10) in the paramagnetic and orbitally ordered phase when approaching the JT transition. Ultimately, in the high-temperature orbitally disordered phase, this excitation becomes heavily overdamped and can be described as broad Debye-like relaxation.

Finally, we also have presented detailed dielectric spectroscopy at audio and radio frequencies. We find a relaxation process with significant dielectric dispersion only in the orbitally ordered and paramagnetic phase. It most likely mirrors the coupled orbital and polar dynamics that causes the ferroelectric order. The relaxation times are in the µs regime and slow down when approaching the structural as well as the magnetic transition. In marked contrast to observations in GaV$_4$S$_8$ [22], there does not seem to be a close connection between this relaxation dynamics and the excitations present in the THz regime.

## ACKNOWLEDGMENTS

We acknowledge partial support by the Deutsche Forschungsgemeinschaft via the collaborative research center TRR 80 "From Electronic Correlations to Functionality" (Augsburg, Munich, Stuttgart).